\documentclass[%
 reprint,
showpacs,preprintnumbers,
 amsmath,amssymb,
 aps,
 pra,
 longbibliography,
 lengthcheck,%
]{revtex4-1}

\usepackage{graphicx,epstopdf, epsfig}
\usepackage{dcolumn}
\usepackage{bm}
\usepackage{hyperref}

\newcommand{\transition}{5s$^2\,^1$S$_0 \to 5$s5p\,$^1$P$_1$\ }
\newcommand{\PRL}{\emph{Phys. Rev. Lett.} }
\newcommand{\jpb}{\emph{J. Phys. B: At. Mol. Opt. Phys.} }

\begin{document}

\title{Spectroscopy of a cold strontium Rydberg gas} 
\author{J. Millen, G. Lochead, G. R. Corbett, R. M. Potvliege and M. P. A. Jones }
\email{m.p.a.jones@durham.ac.uk}
\affiliation{Department of Physics, Durham University, Durham DH1 3LE, United Kingdom}

\date{\today}

\begin{abstract}
 We present a study of a cold strontium Rydberg gas. The narrowband laser excitation of Rydberg states in the range  $n=20-80$ from a 6~mK  cloud of strontium atoms is detected using the spontaneous ionization of the Rydberg atoms. Using a high-resolution step-scanning technique, we perform detailed measurements of the Stark maps of selected Rydberg states. We find excellent agreement between the measured Stark maps and a numerical calculation based on an independent-electron model. Finally we show that excitation of the second valence electron can be used to probe the dynamics of the Rydberg gas  with nanosecond temporal resolution via autoionization.
\end{abstract}
\maketitle

\section*{Introduction}

Rydberg atoms offer an ideal opportunity to study strongly interacting quantum systems, due to the long-range dipole-dipole interactions between them. Two individually trapped neutral atoms excited to Rydberg states have been shown to exhibit a high degree of entanglement, through the dipole blockade mechanism \cite{lukin01, saffman09, gaetan09, heidemann07}. These entangled states are important for the study of quantum information, as they can be used to perform quantum gate operations \cite{saffman10}. In mesoscopic atom clouds, the dipole blockade effect  leads to the formation of highly-correlated many-body states \cite{weimer08} that can exhibit long-range crystalline order \cite{pohl10,lesanovsky10}. The dipole blockade also modifies the atom-light interactions \cite{schempp10}, giving rise to cooperative optical effects \cite{pritchard10} which could be exploited to create photonic phase gates \cite{friedler05}. The long-range interactions cause Rydberg gases to spontaneously evolve into cold neutral plasmas \cite{robinson00} through spontaneous ionization. 

Most experiments in this field have used alkali metal atoms, where only a single valence electron is available for manipulation. Divalent atoms, such as strontium, open up new opportunities for the study of Rydberg gases. For Rydberg states of low angular momentum $L$, excitation of the inner valence electron leads to autoionization, which can be used as a state-selective probe of the interactions in a cold Rydberg gas \cite{millen10}. Off resonance, the strong optical transitions of the ionic core can be exploited to create optical dipole traps for Rydberg atoms \cite{us}. For states of high $L$ the inner valence electron can be excited without ionizing the atom \cite{jones90}, raising the possibility of imaging the Rydberg atoms, in a similar manner to the direct imaging of Sr$^+$ ions in a cold plasma \cite{simien04}.

In this paper we discuss the creation and manipulation of a cold strontium Rydberg gas. We describe the cold Rydberg gas apparatus (section~\ref{expt}), and present a high-resolution spectroscopic study of selected Rydberg states at principal quantum number $n>50$ (section~\ref{spect}). In order to understand the behaviour of cold Rydberg gases, a clear understanding of the nature of the dipole-dipole interactions is essential. This requires a knowledge of the dipole matrix elements. We present a detailed numerical calculation of the Stark maps based on dipole matrix elements  calculated using an independent electron model, and find very good quantitative agreement with experiment (section~\ref{simulation}). Finally we show that laser excitation of the second valence electron can be used as a high time-resolution probe of the Rydberg gas (section~\ref{auto})

\section{Experimental set-up} \label{expt}

\begin{figure}
 \includegraphics{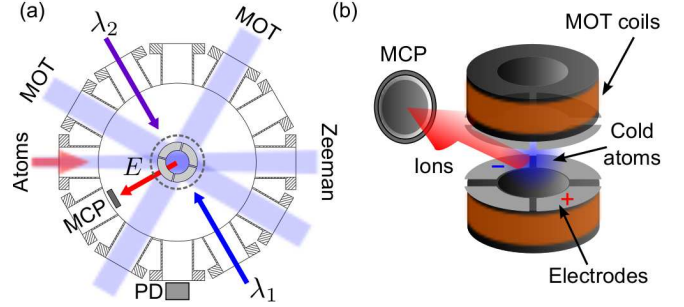} \centering
\caption{\label{electrodes}(a) A scale diagram of the vacuum chamber. Zeeman-slowed atoms enter from the left. Electrodes create an electric field $E$ that directs ions to the  MCP. The fluorescence from the MOT is monitored by a photodiode (PD). The Rydberg excitation beams $\lambda_1$ and $\lambda_2$   (see figure~\ref{timing}(a)) are also shown. (b) A schematic of the apparatus inside the vacuum chamber. The MOT coils are wound on copper formers, which also support  eight electrodes in a split-ring geometry.}
\end{figure}

Cold $^{88}$Sr atoms are prepared in a standard six-beam  magneto-optical trap (MOT) operating on the \transition transition at 461~nm. This light is produced by a frequency doubled diode laser system (Toptica TA-SHG). To stabilize the laser at 461~nm, we perform polarization spectroscopy \cite{javaux10} in a strontium dispenser cell \cite{bridge09}. A schematic of the trapping region is shown in figure~\ref{electrodes}(a). A MOT for strontium requires a magnetic field gradient of $\sim30$~G\,cm$^{-1}$. The field coils are mounted inside the vacuum system, so that this field can be produced without the need for water cooling. The coils are made from Kapton insulated copper wire wound onto copper formers.

A micro-channel plate (MCP) detector is mounted inside the vacuum system to detect ions produced in the Rydberg gas. The MCP is positioned away from the line of-sight of the atomic beam, as shown on figure~\ref{electrodes}(a), and is further protected by a baffle. A wire mesh grid electrode placed in front of the MCP is used to control the stray electric field from the detector. The grid is typically held at $-10$~V. An additional electric field is used to direct ions to the MCP. The field is created by eight independently controllable electrodes, arranged in a split-ring geometry \cite{low07}. The electrodes are mounted onto the MOT coil formers, as shown in figure~\ref{electrodes}(b), and insulated from the copper with ceramic spacers.

The trap is loaded from a Zeeman slowed atomic beam. Strontium is heated in an oven, which consists of heater wire wrapped around the outside of a standard stainless steel vacuum tube \cite{courtillot03}. The hot strontium is collimated using a bundle of 8~mm long, 170~$\mu$m diameter steel capillaries, held in a nozzle which is separately heated to prevent condensation of the strontium. The hot atomic beam passes through a 30~cm long solenoid-type Zeeman slower. The Zeeman slower is encased in a mild steel yoke to boost the magnetic field at the ends of the slower, and further shield the trapped atoms from stray magnetic fields. We typically trap $4 \times 10^6$ atoms, at a density of 2$\times$10$^{10}$~cm$^{-3}$ and a temperature of  6~mK.


The Rydberg sample is prepared using a two-step excitation $\lambda_1+\lambda_2$, as illustrated in figure~\ref{timing}(a). The light for the first step, $\lambda_1 = \mathrm{461}$~nm, is derived from the cooling laser, and is resonant with the \transition~transition. The light for the second step (the Rydberg excitation), $\lambda_2$, is produced by one of two frequency doubled diode laser systems (Toptica DLS SHG), at 420~nm or 413~nm, depending on the range of Rydberg states we wish to access. Both beams are pulsed on simultaneously for 4~$\mu$s. The beams are overlapped, counter-propagating, and linearly polarized in the vertical direction. They propagate normal to the direction of the electric field at the position of the cold atoms. 

The wavelength of the fundamental of the Rydberg excitation laser is measured using a High Finesse WS7 wavemeter, and is converted into the output laser frequency $\omega_2$.

The next section outlines a method for performing Rydberg state spectroscopy using this experimental set-up.

\section{High resolution Rydberg spectroscopy} \label{spect}

We perform spectroscopy by varying the frequency $\omega_2$ of the Rydberg excitation laser, and measuring the amount of spontaneous ionization. Spontaneous ionization has been observed in nearly all experiments on cold Rydberg gases. The dominant mechanism is thought to be interaction-enhanced Penning ionization \cite{li05, amthor09}. In our experiment ionization could also arise from collisions with hot atoms not captured by the Zeeman slower \cite{robinson00}. We observe ionization for all of the states we have studied, which include states of principal quantum number $n = 18$ and above. The ionization is present regardless of whether the interatomic interactions are attractive (as is the case for the $^1$S$_0$ states in Sr \cite{us}), or repulsive (as is the case for the $^1$D$_2$ states above $n = 25$). The ionization is rapid \cite{amthor07}; we observe ionization within 1~$\mu$s of starting the Rydberg excitation.  At the Rydberg densities used in this paper ($< 5\times 10^8$~cm$^{-3}$), the spontaneous ionization signal is proportional to the Rydberg excitation laser intensity. At higher Rydberg densities we observe the formation of a cold plasma \cite{millen10}, which leads to avalanche ionization \cite{robinson00}. 

The spontaneously created ions are directed to the MCP with a 4~$\mu$s long electric field pulse. The amplitude of this pulse (4~V\,cm$^{-1}$) is not sufficient to field ionize any of the Rydberg states described in this paper. The signal from the MCP is amplified, recorded on a digital oscilloscope and integrated to yield a total ion signal in V\,$\mu$s.

\begin{figure}
\includegraphics{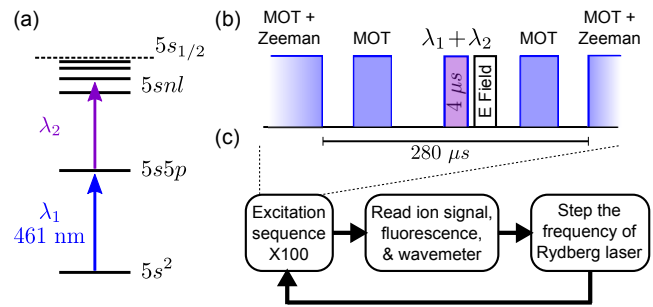} \centering
\caption{\label{timing}(a) Energy level diagram for the two-step excitation, $\lambda_1 + \lambda_2$, to a Rydberg state 5s$nl$. (b) Timing diagram for the excitation sequence. The Rydberg state is populated after the ground-state atoms are released from the MOT. The MOT light is pulsed before and after the Rydberg excitation. (c) Experimental sequence for the step-scan technique. The Rydberg excitation sequence is repeated 100 times. The averaged ion signal and fluorescence, and the frequency $\omega_2$ of $\lambda_2$, are recorded. The frequency $\omega_2$ is stepped, and the process repeated to produce a spectrum.}
\end{figure}

The sequence that we use to perform spectroscopy is illustrated in figure~\ref{timing}(b). A pulse of MOT light before and after the Rydberg excitation is used to measure the ground state atom number and any loss of atoms. This excitation sequence is repeated 100 times, and the ionization and fluorescence signals are averaged on the oscilloscope. After the 100 repeats the average signals are recorded, and the frequency $\omega_2$ is sampled 100 times. A voltage is applied to  the laser piezo to step the frequency $\omega_2$, and whole process is repeated, as illustrated in figure~\ref{timing}(c). By repeatedly stepping $\omega_2$ we can build up a Rydberg state spectrum. We refer to this method as the ``step-scan'' technique. The entire stepping and recording process is automated under computer control.

\begin{figure}
 \includegraphics{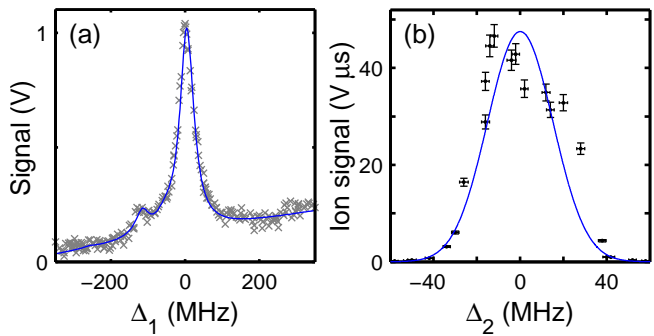} \centering
\caption{\label{stepscan} Example spectra obtained using the step-scan technique. The detunings $\Delta_1,\Delta_2$ are defined relative to our measured line centres. (a) Saturated absorption spectrum of the \transition~transition. The fit is a sum of six Lorentzians.  (b) Spectrum of the 5s19d\,$^1$D$_2$ state. The solid line is a Gaussian fit with a FWHM of 48~MHz, which matches the linewidth of the 5s5p\,$^1$P$_1$ state (32~MHz) with expected power-broadening. }
\end{figure}

Importantly, measuring $\omega_2$ at each step enables us to negate the effect of slow drifts in the laser frequency. To characterize the accuracy of this step-scan technique, we performed saturated absorption spectroscopy of the \transition~transition, as shown in figure~\ref{stepscan}(a). The frequency $\omega_1$ of the 461~nm laser is stepped across the resonance in steps of 1.4~MHz, which are clearly resolved by the wavemeter. We fit the spectrum with a sum of six Lorentzians, one for each isotope and hyperfine component \cite{mauger07}. The free parameters are the overall amplitude, a sloping background and a frequency axis scaling parameter $\alpha$, such that $\omega_1 \to \alpha \omega_1$. From the fit, we obtain $\alpha = \mathrm{1.03}\pm\mathrm{0.02}$, indicating that the relative frequency accuracy is very good. By comparing our measured wavelengths to literature values, we estimate that the absolute accuracy of the wavemeter is $\sim 200$~MHz. This is consistent with the wavemeter specification when using multi-mode optical fibres.

\begin{figure}
 \includegraphics{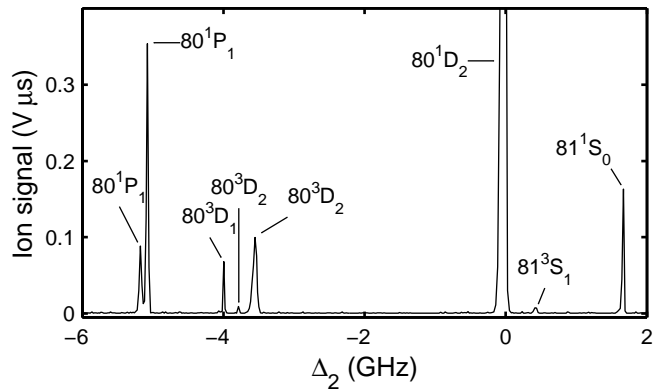} \centering
\caption{\label{n80}Spectrum in the region $n = \mathrm{80}$, in the presence of a 0.35~V\,cm$^{-1}$ electric field. The detuning $\Delta_2$ is relative to the energy of the $80\,^1$D$_2$ state in zero field. The $80\,^1$D$_2$ state signal peaks at 1.4~V\,$\mu$s. The electric field allows otherwise forbidden transitions, and has split the $80\,^1$P$_1$ and $80\,^3$D$_2$ states into their $\vert m_J \vert$ components.  The lines are assigned using the single electron simulation. }
\end{figure}

Examples of the Rydberg spectra that we obtain are shown in figures \ref{stepscan}(b) and \ref{n80}. Figure \ref{stepscan}(b) shows a high-resolution scan across a single Rydberg state. The linewidth of the Rydberg excitation is consistent with the power-broadened linewidth of the intermediate 5s5p\,$^1$P$_1$ state. In figure~\ref{n80} the Rydberg laser is stepped over 8~GHz, in steps of 7~MHz, in the vicinity of the 5s80d\,$^1$D$_2$ state. This spectrum was acquired in the presence of a constant electric field of 0.35~V\,cm$^{-1}$, which allows the excitation of the dipole-forbidden $^1\mathrm{P}_1$ state. Triplet states are also visible in the spectrum, indicating that singlet-triplet mixing \cite{esherick77} is still present even at $n=80$. The 81\,$^3\mathrm{S}_1$ peak is $\sim200$ times smaller than the principal 80\,$^1\mathrm{D}_2$ peak, which highlights that spontaneous ionization is detectable even for these extremely weakly populated states.

\section{Single electron model and Stark maps} \label{simulation}

Rydberg atoms exhibit greatly enhanced interatomic interactions as compared to ground state atoms. To calculate the size and form of these interactions the dipole matrix elements between different atomic states must be known. This requires a knowledge of the atomic state wavefunctions. For the alkali elements, the wavefunctions are commonly calculated by considering a single electron moving in a model potential, which incorporates the effect of the valence electron penetrating the closed electronic shells \cite{theodosiou84}.

The situation for alkali earth elements is more complex, since there are two valence electrons. This leads to perturbations in the energy level structure due to the presence of doubly excited perturber states \cite{bookoftom}. To fully describe the energy level structure of the alkali earth elements, multi-channel quantum defect theory (MQDT) must be used \cite{aymar96}. However, \cite{chan01} recently obtained good agreement between experimental and theoretical Sr Stark maps in the region of principal quantum number $n = 12$, using dipole matrix elements obtained by treating the two electrons as completely independent and moving in a simple model potential. In the following, we extend this model to much higher values of $n$ and total orbital angular momentum $L$, and find that it remains in good agreement with our experimentally measured Stark maps. 

Our simulation uses known quantum defects \cite{beigang82, esherick77, spectro}, and extrapolates them to regions where they have not been previously measured. Using these values we fit a model potential of the Klapisch form \cite{klapisch71} using a simulated annealing algorithm. Radial wavefunctions are calculated, using the Numerov method \cite{hajj80}, by considering a single electron moving in this model potential. Dipole matrix elements are calculated from the radial wavefunctions \cite{potvliege98}. To calculate a Stark map in the region $n = n_0$ states of $n = n_0 \pm 5$ are included, and the angular momentum $L$ states are truncated at $L= 40$.

\begin{figure}
 \includegraphics{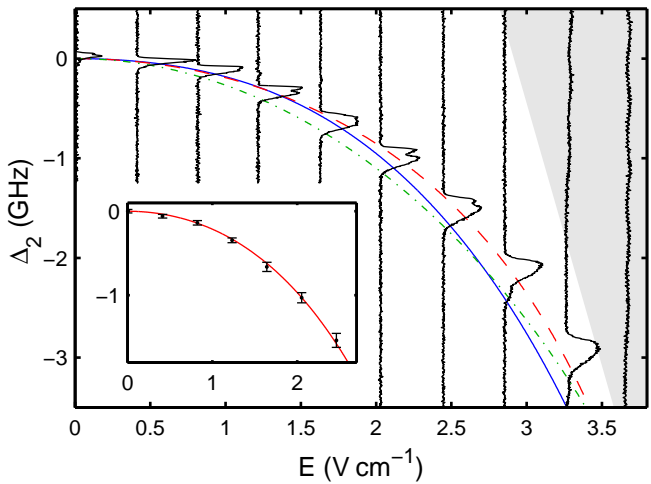} \centering
\caption{\label{stark}Stark map in the region of the 5s56d\,$^1$D$_2$ state. The detuning $\Delta_2$ is relative to the energy of the state in zero electric field. The blue (solid), red (dashed), green (dot-dashed) lines are the simulated $\vert m_J\vert= 0, 1, 2$ components respectively. The grey region is the high angular momentum manifold. The inset shows the fit of the simulated Stark shift averaged over the three $\vert m_J \vert$ components (red line) to the experimental centre frequency (black dots), extracted using a Gaussian fit to the lineshape at each electric field. The linewidth (FWHM) is indicated by the error bars.}
\end{figure}

\begin{figure}
 \includegraphics{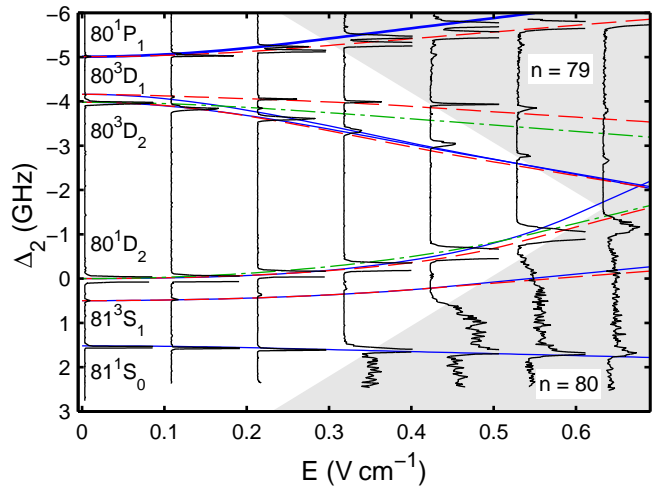} \centering
\caption{\label{stark80}A Stark map in the region of the 5s80d\,$^1$D$_2$ state.  Solid black lines are the experimental spectra. The detuning $\Delta_2$ is relative to the energy of the $80\,^1$D$_2$ state in zero electric field. The blue (solid), red (dashed), green (dot-dashed) lines are the simulated $\vert m_J\vert= 0, 1, 2$ components respectively, and the grey shaded areas represent the simulated high angular momentum manifolds. The $80\,^3$D$_3$  state is doubly forbidden and is not observed. }
\end{figure}

The Stark map is measured using the step-scan technique, with a static electric field applied during the Rydberg excitation. Measured and simulated Stark maps are shown in the region of the 5s56d\,$^1$D$_2$ state in figure~\ref{stark}, and the 5s80d\,$^1$D$_2$ in figure~\ref{stark80}. In order to compare experiment and theory, the calibration factor between the applied voltage and the electric field must be determined. The calibration factor is obtained by fitting the simulated Stark map to the experimental data, as shown in the inset to figure~\ref{stark}. There are three fitting parameters: the calibration factor, the voltage offset due to the stray electric field, and a frequency offset due to the wavemeter uncertainty. We constrain the voltage offset using the variation in the amplitude of the 80\,$^1$P$_1$ peak with electric field.  This state cannot be populated in zero field, since we excite via the 5s5p\,$^1$P$_1$ state. The fit shown in the inset to figure~\ref{stark} yields a calibration factor of $0.205\pm0.005$ V\,cm$^{-1}$ per volt applied. A similar fit to the $n=80$ Stark map yields a calibration factor of $0.215\pm0.005$ V\,cm$^{-1}$ per volt applied. The corresponding value for the stray electric field obtained from the voltage offset is $2.8 \pm 0.4$~mV\,cm$^{-1}$. This is consistent with the estimated electric field due to the MCP grid.

These calibration curves highlight the excellent quantitative agreement between the single electron model and the experiment for the $^1$D$_2$ states. In figure~\ref{stark}(a), the splitting between the $\vert m_J\vert = 0, 1, 2$ (blue solid, red dashed, green dot-dashed lines) agree well with the data (black lines), and the interaction with the higher angular momentum manifold (grey region) is correctly predicted and visible in the data. The agreement is also very good across  the $n = 80$ Stark map shown in figure~\ref{stark80}. The only previously measured state energy in this region is for the 5s80d\,$^1$D$_2$ state \cite{beigang82}, and we have adjusted the quantum defects for the other series to match our measured splitting from the $80\,^1$D$_2$ state in zero field. This requires a change of $<1$\% from the extrapolated quantum defects. The interaction of the low $L$ states with the high angular momentum manifolds (shaded grey areas on figure~\ref{stark80}) is clearly visible. The agreement is better for the singlet states than the triplet states. More quantum defects must be found by extrapolation for the triplet series, as the spectroscopic data is less complete. Alternatively, the disagreement may reflect the fact that the single electron model ignores the mixing between the singlet and triplet states.

We have shown that a single electron model can be used to calculate dipole matrix elements for strontium, over a large range of $n$. The single electron model is therefore appropriate for calculating the dipole-dipole interactions, which depend on these matrix elements \cite{us}. The single electron model does not work for predicting state lifetimes and line strengths, which are strongly perturbed by the presence of doubly excited states. To fully describe the energy level structure of the alkali earth elements multi-channel quantum defect theory (MQDT) must be used \cite{aymar96}, which takes into account the divalent nature of the atom.

\section{Pulsed autoionization} \label{auto}

Rather than relying on spontaneous ionization, we can detect the Rydberg atoms using autoionization by exciting the inner valence electron. If an atom is initially prepared in a Rydberg state, the inner valence electron can be excited independently, a process known as isolated core excitation \cite{cooke78}. The doubly excited states couple not only to each other (as with singly excited states), but also to the degenerate continuum. For states of low angular momentum $L$ this leads to rapid ionization, and these states are referred to as ``autoionizing''. There has been much research into the nature of autoionizing resonances, with MQDT proving to be extremely powerful for describing doubly excited states \cite{aymar96}. The autoionization process has a large cross-section, and hence produces significantly more ions than spontaneous ionization. We typically find that by exciting an autoionizing resonance the signal is increased by a factor of 100.

We prepare atoms in the 5s$nl$ Rydberg state, using the same sequence as before (figure~\ref{timing}). At a variable time $\Delta$t after the end of the excitation to the Rydberg state, a 10~ns light pulse  at $\lambda_3 = 408$~nm excites the inner valence electron, creating the autoionizing state 5p$_{3/2}n'l$. The light $\lambda_3$ is provided by a tunable pulsed dye laser system, with a bandwidth of a few GHz, and a repetition rate of 10~Hz. The autoionizing pulse is immediately followed by an electric field pulse that directs the ions to the MCP. 

At each value of the wavelength $\lambda_3$ and delay $\Delta$t we measure a Rydberg spectrum using the step-scan technique described previously. The spectrum is normalized by the average pulse intensity measured on a photodiode, and the ground state atom number. From this spectrum we extract the peak autoionization-enhanced ion signal, S, in units of V\,$\mu$s\,$\mu$W$^{-1}$\,cm${^2}$ per 10$^6$ atoms.

\begin{figure}
 \includegraphics{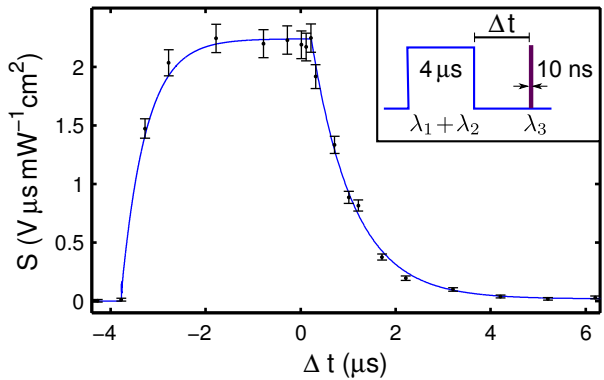} \centering
\caption{\label{lowntime}The autoionization signal for the 5s19d\,$^1$D$_2$ state as a function of time. $\Delta \mathrm{t} = 0$ is defined as the end of the 4~$\mu$s long excitation to the Rydberg state $\lambda_1 + \lambda_2$, as illustrated in the inset. The solid line is the solution to the optical Bloch equations.}
\end{figure}

By fixing the wavelength $\lambda_3$ and scanning the delay  $\Delta$t across the 4~$\mu$s Rydberg excitation pulse,  the evolution of the Rydberg population can be mapped with high time resolution as shown in figure~\ref{lowntime}. Both the Rydberg excitation and subsequent decay are clearly visible. By fitting an exponential to the decay we find a state lifetime of $\tau_{19\mathrm{D}} = 880 \pm 30$~ns. The complete time evolution can be modelled by solving the three level optical Bloch equations for the scheme 5s$^2\,^1$S$_0 \to 5$s5p\,$^1$P$_1 \to 5$s19d\,$^1$D$_2$. 
The Rabi frequencies and decay constants for each step of the excitation are constrained using our measurements of the laser intensity and the lifetime $\tau_{19\mathrm{D}}$, and the oscillator strength for the  5s5p\,$^1$P$_1 \to 5$s19d\,$^1$D$_2$ transition from \cite{haq07}. The only remaining free parameter is the amplitude scale. The fit is in very good agreement with the data, as can be seen in figure~\ref{lowntime}. This indicates that the autoionization signal is proportional to the Rydberg state population. An identical analysis for the 5s20s\,$^1$S$_0$ state yields a state lifetime of 3.1$\pm$0.1~$\mu$s.

\begin{figure}
 \includegraphics{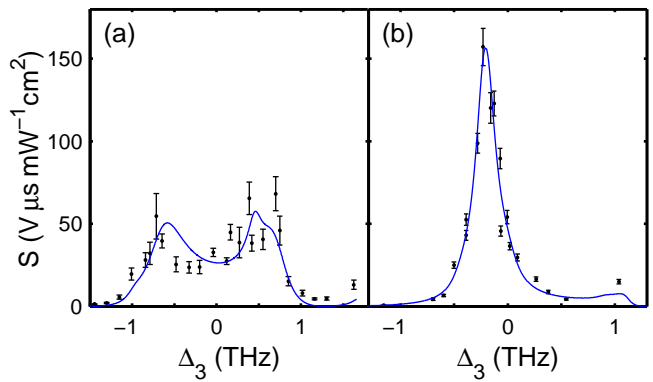} \centering
\caption{\label{lownspect}Autoionization spectra of the (a) 5s19d\,$^1$D$_2$ and (b) 5s20s\,$^1$S$_0$ states. The spectra are fit with a six-channel MQDT model. $\Delta_3$ is the detuning of $\omega_3$ from the 5s$_{1/2} \to 5$p$_{3/2}$ ion transition. These spectra are taken at a delay $\Delta \mathrm{t} = \mathrm{0.5}~\mu$s.}
\end{figure}

By fixing the delay $\Delta$t, and varying the wavelength $\lambda_3$, the spectrum of the  5s$nl \to 5$p$_{3/2}n'l$ autoionizing transition can be measured. MQDT allows us to reproduce the spectra of the autoionizing states, which in turn allows us to identify exactly which state a sample is in from its autoionization spectrum. The spectra are fit with a six-channel MQDT model, using the same fitting parameters as \cite{xu87} for the $19\,^1$D$_2$ state (with the elliptical polarization of $\lambda_3$ included in the model), and \cite{xu86} for the $20\,^1$S$_0$ state. The difference in the shape of the spectra is due to the difference $\epsilon = \delta_b-\delta_a$ between the bound (Rydberg) state quantum defect $\delta_b$, and the autoionizing state quantum defect $\delta_a$. For $\epsilon \simeq \mathrm{0}$ (modulo-one) each Rydberg state only overlaps with a single autoionizing state, and the autoionizing spectrum is a single peak, as seen for the $20\,^1$S$_0$ state with $\epsilon = \mathrm{0.15}$. For $\epsilon \simeq \mathrm{0.5}$ (modulo-one) each Rydberg state overlaps with two autoionizing states, and the autoionizing spectrum is double-peaked, as seen for the $19\,^1$D$_2$ state with $\epsilon = \mathrm{0.46}$

The spectra in figure \ref{lownspect} illustrate that autoionization can be used for state-selective detection. It is particularly sensitive to changes in  $L$.  In previous work we have  identified, and quantified, transfer of population from one Rydberg state to another through analysis of the autoionization spectra \cite{millen10}. In the region $n\approx20$  the autoionization spectra are approximately 1~THz wide as shown in figure~\ref{lownspect}. Even at $n = 56$, the width of the autoionizing spectrum for low-$L$ states is $>10$~GHz  \cite{millen10}. The high temporal resolution of the autoionization technique displayed in figure \ref{lowntime} can therefore be fully exploited without compromising its state-selectivity. The combination of high temporal resolution, state-selectivity and large signal-to-noise ratio make autoionization a powerful technique for probing the dynamics of cold Rydberg gases.

\section*{Conclusion}

The use of alkali earth elements such as strontium offer new ways to study and manipulate Rydberg atoms. We have performed the first study of Rydberg states in a cold gas of strontium. We have used, and experimentally verified, a simple single-electron model to find dipole matrix elements, which can be used to calculate interatomic interaction strengths. We have also used autoionization through two-electron excitation as a high-yield probe of a cold Rydberg gas, which can measure state population dynamics with high temporal resolution. By focusing the autoionizing laser this technique could be used to spatially probe the Rydberg gas, on length scales comparable to the dipole blockade radius.

We thank J.D. Pritchard for assistance with the dipole matrix element calculations, and C. S. Adams and D. Carty for the loan of equipment. This work was supported by EPSRC Grants No. EP/D070287/1 and No. LP/82000, and by Durham University.


\section*{Bibliography}

\begin{thebibliography}{99}
\bibitem{lukin01}Lukin~M~D, Fleischhauer~M and Cote~R 2001 \PRL {\bf 87} 037901

\bibitem{saffman09}Urban~E, Johnson~T~A, Henage~T, Isenhower~L, Yavuz~D~D, Walker~T~G and Saffman~M 2009 \emph{Nature Physics} {\bf 5} 110--114

\bibitem{gaetan09}Ga{\"e}tan~A, Miroshnychenko~Y, Wilk~T, Chotia~A, Viteau~M, Comparat~D, Pillet~P, Browaeys~A and Grangier~P 2009 \emph{Nature Physics} {\bf 5} 115--118

\bibitem{heidemann07}Heidemann~R, Raitzsch~U, Bendkowsky~V, B{\"u}tscher~B, L{\"o}w~R, Santos~L and Pfau~T 2007 \PRL {\bf 99} 163601

\bibitem{saffman10}Saffman~M, Walker~T~G and M{\o}lmer~K 2010 \emph{Rev. Mod. Phys.} {\bf 82} 2313--2363

\bibitem{weimer08}Weimer~H, L\"{o}w~R, Pfau~T and B\"{u}chler~H~P 2008 \PRL {\bf 101} 250601 

\bibitem{pohl10}Pohl~T, Demler~E and Lukin~M~D 2010 \PRL {\bf 104} 043002

\bibitem{lesanovsky10}Schachenmayer~J, Lesanovsky~I, Micheli~A and Daley A~J 2010 \emph{New J. Phys.} {\bf 12} 103044 




\bibitem{schempp10}Schempp~H, G{\"u}nter~G, Hofmann~C~S, Giese~C, Saliba~S~D, DePaola~B~D, Amthor~T and Weidem{\"u}ller~M 2010 \PRL {\bf 104} 173602

\bibitem{pritchard10}Pritchard~J~D, Maxwell~D, Gauget~A, Weatherill~K~J, Jones~M~P~A and Adams~C~S 2010 \PRL {\bf 105} 193603

\bibitem{friedler05}Friedler~I, Petrosyan~D, Fleischhauer~M and Kurizki~G 2005 \emph{Phys. Rev. A} {\bf 72} 043803

\bibitem{robinson00}Robinson~M~P, {Laburthe Tolra}~B, Noel~M~W, Gallagher~T~F, Pillet~P 2000 \PRL {\bf 85} 4466--4469

\bibitem{millen10}Millen~J, Lochead~G and Jones~M~P~A 2010 \PRL {\bf 105} 213004

\bibitem{us}Mukherjee~R, Millen~J, Nath~R, Jones~M~P~A and Pohl~T \jpb {\bf ?} ?

\bibitem{jones90}Jones~R~R, Dai~C~J and Gallagher~T~F 1990 \emph{Phys. Rev. A} {\bf 41} 316--326

\bibitem{simien04}Simien~C~E, Chen~Y~C, Gupta~P, Laha~S, Martinez~Y~N, Mickelson~P~G, Nagel~S~B and Killian~T~C 2004 \PRL {\bf 92} 143001


\bibitem{javaux10}Javaux~C, Hughes~I~G, Lochead~G, Millen~J and Jones~M~P~A 2010 \emph{Eur. Phys. J. D} {\bf 57} 151--154

\bibitem{bridge09}Bridge~E~M, Millen~J, Adams~C~S and Jones~M~P~A 2009 \emph{Rev. Sci. Instr.} {\bf 80}  013101

\bibitem{low07}L{\"o}w~R, Raitzsch~U, Heidemann~R, Bendkowsky~V, Butscher~B, Grabowski~A and Pfau~T 2007 \emph{Preprint} quant-ph/0706.2639

\bibitem{courtillot03}Courtillot~I, Quessada~A, Kovacich~R~P, Zondy~J-J, Landragin~A, Clairon~A and Lemonde~P 2003 \emph{Optics Letters} {\bf 28} 468--470




\bibitem{li05}Li~W, Tanner~P~J and Gallagher~T~F 2005 \PRL {\bf 94} 173001

\bibitem{amthor09}Amthor~T, Denskat~J, Giese~C, Bezuglov~N~N, Ekers~A, Cederbaum~L~S and Weidem{\"u}ller~M 2009 \emph{Eur. Phys. J. D} {\bf 53} 329--335

\bibitem{amthor07}Amthor~T, Reetz-Lamour~M, Westermann~S, Denskat and Weidem{\"u}ller~M 2007 \PRL {\bf 98} 023004

\bibitem{mauger07}Mauger~S, Millen~J and Jones~M~P~A 2007 \jpb {\bf 40} F319--325


\bibitem{esherick77}Esherick~P 1977 \emph{Phys. Rev. A} {\bf 15} 1920--1936

\bibitem{theodosiou84}Theodosiou~C~E 1984 \emph{Phys. Rev. A} {\bf 230} 2881--2909

\bibitem{bookoftom}Gallagher~T~F 1994 \emph{Rydberg Atoms} (Cambridge University Press)

\bibitem{aymar96}Aymar~M, Greene~C~H and {Luc-Koenig}~E 1996 \emph{Rev. Mod. Phys} {\bf 68} 1015--1123

\bibitem{chan01}Zhi~M~C, Dai~C~J and Li~S~B 2001 \emph{Chinese Physics} {\bf 10} 929--934


\bibitem{beigang82}Beigang~R, L\"ucke~K, Timmermann~A, West~P~J and Fr\"olich~D 1982 \emph{Opt. Commun.} {\bf 42} 19--24

\bibitem{spectro}Moore~C~E 1952 \emph{Atomic Energy Levels (Chromium Through Niobium)} (US government printing office, Washington) Vol.~2; \newline Garton~W~R~S and Codling~K 1968 \jpb {\bf 1} 106--113;  \newline Rubbmark~J~R and Borgstr{\"o}m~S~A 1978 \emph{Physica Scripta} {\bf 18} 196--208; \newline Armstrong~J~A, Wynne~J~J and Esherick~P 1979 \emph{J. Opt. Soc. Am.} {\bf 69} 211--230; \newline Beigang~R, L\"ucke~K, Schmidt~D, Timmermann~A and West~P~J 1982 \emph{Physica Scripta} {\bf 26} 183--188; \newline Beigang~R and Schmidt~D 1983 \emph{Physica Scripta} {\bf 27} 172--174; \newline Dai~C~J 1995 \emph{Phys. Rev. A} {\bf 52} 4416--4424

\bibitem{klapisch71}Klapisch M 1971 \emph{Comput. Phys. Comm.} {\bf 2} 239--260

\bibitem{hajj80}Hajj~F~Y 1980 \jpb {\bf 13} 4521--4528

\bibitem{potvliege98}Potvliege R M 1998 \emph{Comput. Phys. Comm.} {\bf 114} 42--93



\bibitem{cooke78}Cooke~W~E, Gallagher~T~F, Edelstein~S~A and Hill~R~M 1978 \PRL {\bf 40} 178--181

\bibitem{haq07}Haq~S-U, Mahmood~S, Kalyar~M~A, Rafiq~M, Ali~R and Baig~M~A 2007 \emph{Eur. Phys. J. D} {\bf 44} 439--447

\bibitem{xu87}Xu~E~Y, Zhu~Y, Mullins~O~C and Gallagher~T~F 1987 \emph{Phys. Rev. A} {\bf 35} 1138--1148

\bibitem{xu86}Xu~E~Y, Zhu~Y, Mullins~O~C and Gallagher~T~F 1986 \emph{Phys. Rev. A} {\bf 33} 2401--2409



\end{thebibliography}
\end{document}